\documentclass[aps,showpacs,twocolumn,superscriptaddress]{revtex4}
\usepackage[utf8x]{inputenc}
\usepackage{ucs}
\usepackage{amsmath}
\usepackage{amsfonts}
\usepackage{amssymb}
\usepackage{appendix}
\usepackage{hyperref}
\usepackage{makeidx}
\usepackage{graphicx}
\usepackage{amsthm}
\usepackage{color}
\usepackage{comment}
\usepackage{rotating}
\usepackage{pictex}
\usepackage{makeidx}
\usepackage{float}
\usepackage{wrapfig}
\usepackage{epsfig}

\newtheorem*{thm*}{Theorem}

\begin{document}
\title{Two-dimensional confined hydrogen: An entropy and complexity approach}
\author{C. R. Estañón}
\email[]{carloscbiuam1@gmail.com}
\affiliation{Departamento de  F\'isica, Universidad Aut\'onoma Metropolitana-Iztapalapa,\\
Av. San Rafael Atlixco 186, Col. Vicentina, CP 09340 CDMX, M\'exico}

\author{N. Aquino}
\email[]{naa@xanum.uam.mx}
\affiliation{Departamento de  F\'isica, Universidad Aut\'onoma Metropolitana de Iztapalapa, CDMX, M\'exico}

\author{D. Puertas-Centeno}
\email[]{david.puertas@urjc.es}
\affiliation{%
	Departamento de Matem\'atica Aplicada, Universidad Rey Juan Carlos, 28933 Madrid, Spain
}
\author{J.S. Dehesa}
\email[]{dehesa@ugr.es}
\affiliation{Departamento de F\'{\i}sica At\'{o}mica, Molecular y Nuclear, Universidad de Granada, Granada 18071, Spain}
\affiliation{Instituto Carlos I de F\'{\i}sica Te\'orica y Computacional, Universidad de Granada, Granada 18071, Spain}

\begin{abstract}

The position and momentum spreading of the electron distribution of the two-dimensional confined  hydrogenic atom, which is a basic prototype of the general multidimensional confined quantum systems, is numerically studied in terms of the confinement radius for the 1s, 2s, 2p and 3d quantum states by means of the main entropy and complexity information-theoretic measures. First, the Shannon entropy and the Fisher information as well as the associated uncertainty relations are computed and discussed. Then, the Fisher-Shannon, LMC and LMC-R\'enyi complexity measures are examined and mutually compared. We have found that these entropy and complexity quantities reflect the rich properties of the electron confinement extent in the two conjugated spaces.

\end{abstract}



\maketitle

\section{Introduction}

The fundamental and practical relevance of the spherically-confined quantum systems has been manifested from the early days of quantum physics  \cite{Michels1937,Sommerfeld1938} up until now \cite{Jaskolski1996,Sabin2009,Sen2009,Sen2014,LeyKoo2018}. They have been used as prototypes to explain numerous phenomena and systems not only in the three-dimensional world but also for non-relativistic and relativistic $D$-dimensional ($D\geq 2$) chemistry and physics \cite{Herscbach1993,Dehesa2011,Rivera2019,Cox1985,Jakubith1985,Beta1985,Naber2019}. For example, pioneered by Gerhard Ertl, the 2007 Nobel Laureate in Chemistry, and his collaborators, surface chemistry has been extensively studied and become a key branch in chemistry  \cite{Cox1985,Jakubith1985,Beta1985}. In physics of materials two-dimensional systems have been used to gain insight into the properties of semiconductors (see e.g., \cite{Levine1965,Li2007}) and they start to play a fundamental role to bring strong and new forms of control to the atomic-scale limit over the dynamics of matter in the extreme confinement of electromagnetic energy by phonon polaritonics \cite{Rivera2019}. Moreover, the properties of the real fluids can be studied by means of cristaline fluids (i.e., with a symmetry) with non-standard dimensions (see e.g., \cite{Costigliola2016}). \\

The idea of two- and multidimensional spherical confinement of atoms has been used not only to simulate the effect of high pressure on the static dipole polarizability in hydrogen \cite{Miller1977} but also to model a great deal of nanotechnological objects such as quantum dots, quantum wells and quantum wires \cite{Harrison2005,Munjal2018}, atoms and molecules embedded in nanocavities as for example in fullerenes, zeolites cages and helium droplets \cite{Sabin2009,Al-Hashimi2012,Sen2014,Aquino2014,Rojas2018,LeyKoo2018}, dilute bosonic and fermionic systems in magnetic traps of extremely low temperatures \cite{Gleisberg2000,Anglin2002,DeMarco1999} and a variety of quantum-information elements \cite{Lukin2018,Yoder2019}. This has provoked a fast development of a density functional theory of independent particles moving in multidimensional central potentials with various analytical forms (see e.g.,\cite{Garza2005,Dehesa2007,Sen2009,Sen2014,LeyKoo2018,Mukherjee2019}).  \\

Most efforts have been centered around the spectroscopic properties and some density-functional descriptors of physical and chemical quantities for the ground state of spherically confined atoms \cite{Sabin2009,Sen2014,LeyKoo2018}. However, not so much is known about the information-theoretic measures of the multidimensional confined systems except for a few recent entropy-like \cite{Sen2005,Aquino2013, Nascimento2018,Jiao2017,Mukherjee2018,Mukherjee2018b,Mukherjee2018a,Munjal2018,Garza2019,Wu2020} and complexity-like \cite{Aquino2013,Majumdar2017} results of the three-dimensional confined hydrogenic atom. The aim of this work is to cover this informational lack by means of the determination of the confinement dependence of some entropy (Shannon, Fisher) and complexity (Fisher-Shannon, LMC and LMC-Rényi) measures for the 1s, 2s, 2p and 3d quantum states of the two-dimensional confined hydrogenic atom (2D-CHA, in short)\cite{Yang1991,Aquino1998,Aquino1998a,Aquino2005,Chaos2005} in both position and momentum spaces. This model has been extensively used to interpret numerous phenomena in quantum chemistry, theory of materials, nanotechnology and quantum information and computation as already mentioned, among other fields. These quantities of entropy and complexity character measure the spatial electron delocalization of the system.\\

The Shannon and Fisher information entropies of a multidimensional quantum state in position space are integral functionals of the probability density $\rho(\vec{r})$ of the state \cite{Shannon1948,Fisher1925,Dehesa2011}. The Shannon entropy, which is a logarithmic functional of $\rho(\vec{r})$, quantifies the total extent of the density. The Fisher information, which is a gradient functional of $\rho(\vec{r})$, measures the pointwise concentration of the electronic probability cloud all over the region wherein the electron moves. These entropic quantities are (i) closely related to fundamental and/or experimentally measurable quantities of electronic systems, (ii) satisfy sharp uncertainty relationships \cite{Bialynicki1975,Beckner1975,Sanchez-Moreno2011}, (iii) the basic variables of two extremization procedures (the maximum entropy method and the principle of extreme physical information, respectively), (iv) indicators of the most distinctive nonlinear phenomena (avoided crossings) encountered in atomic and molecular spectra under external fields \cite{Gonzalez2003,Gonzalez2005}, and (v) identifiers of the shape effect of quantum heterostructures \cite{Munjal2018}, among many other properties. \\

The Fisher-Shannon, LMC and LMC-Rényi complexities of the system \cite{Romera2004,Angulo2008,Catalan2002,Lopez-Ruiz2009,Sanchez-Moreno2014} are spreading quantities composed by two entropic measures of global and local character in the Fisher-Shannon case, and by two global entropic measures in the other two cases. Each of these intrinsic complexity measures quantifies the combined balance of two macroscopic facets of the multidimensional quantum probability density of the system. They have a number of interesting properties: vanishing for the two extreme probability densities which corresponds to perfect order and maximum disorder \cite{Sanchez-Moreno2014}, invariance under translation, scaling and replication transformations \cite{Lopez-Ruiz2011}, monotonicity \cite{Rudnicki2016, Puertas-Centeno2018}, and others \cite{Yamano2004,Yamano2004b}. These statistical complexity measures have been widely applied to gain insight into the internal structure of atomic and molecular systems \cite{Dehesa2011,Nagy2009,Antolin2009,Dehesa2013,Guerrero2011,Borgoo2011}, electron correlation \cite{Romera2004}, topological quantum phase transitions \cite{Bolivar2018} and macroscopic features of biological and pharmacological molecules (e.g., amino acids, sulfanamides, ...) \cite{Esquivel2016,Lopezrosa2016}, among others.\\


The structure of the work is the following. In Section \ref{Methodology} we briefly describe the quantum probability density for the stationary states of  the two-dimensional confined hydrogen atom. In Section \ref{Entropic} the Shannon and Rényi entropies and the Fisher information of the 2D-CHA are analyzed. In Section \ref{Complexity}  the behavior of the Fisher-Shannon, LMC and LMC-Rényi complexity measures is studied and discussed. Finally, in Section \ref{Conclusions} some conclusions and open problems are given.

\section{Two-Dimensional confined Hydrogen atom}\label{Methodology}

In this Section we gather the electronic probability densities of the quantum stationary states $(n, m)$ of the two-dimensional confined hydrogen atom (i.e., an electron moving around the nucleus in a circular region of radius $r_{0}$  with impenetrable walls and the nucleus located at the center) in both position and momentum spaces. Atomic units are used throughout the paper.\\

\subsection{Wavefunctions}

The time-independent Schr\"{o}dinger equation of a two-dimensional hydrogen atom is given by
\begin{equation}\label{eqI_cap1:ec_schrodinger}
\left( -\frac{1}{2} \vec{\nabla}^{2}_{2} - \frac{1}{r}
\right) \Psi \left( \vec{r} \right) = E \Psi \left(\vec{r} \right),
\end{equation}

where $\vec{\nabla}_{2}$ denotes the two-dimensional gradient operator and the  electronic position vector $\vec{r}  =  (x_1 ,x_2)$ in polar coordinates is  given as $(r,\theta)$, where $r \equiv |\vec{r}| = \sqrt{x_1^2 + x_2^2}\in [0 , r_{0}]$  
and with $\theta\in [0 , 2\pi)$. It is known that the radial symmetry of the system implies that the wavefunction is separable in the radial and angular parts as $\Psi(\vec r)=R(r) \phi (\theta)$. Note also that with $r_{0}\rightarrow \infty$ one has the free (unconfined) case; then, the radial wavefunction has the expression \cite{Yang1991, Aquino1998}
\begin{equation}
R_{n,m}(r)=N_{n,m}e^{-\beta_n r/2} \left( \beta_n r\right)^{\lvert m\rvert}L_{n-\lvert m\rvert-1}^{2\lvert m\rvert}\left(\beta_n r\right),
\end{equation}
where $\beta_n=\frac{2\,Z}{n-1/2}$, $N_{n,m}$ denotes the  normalization constant 
\begin{equation} \label{constant}
N_{n,m}=\frac{\beta_n}{\sqrt{2n\,(n-m)_{2m-1}}},
\end{equation}
and the angular part is given by $ \phi_m(\theta)=\frac{e^{im\theta}}{\sqrt{2\pi}}.$ The quantum numbers $(n, m)$ which characterize the state have the values $n=1,2,\dots$ and $m=0,1,\dots,n-1$. The states with $m=n-1$ are usually called by circular states.

To find the corresponding wavefunctions in the two-dimensional confined system, we have followed the variational methodology used by Aquino and Casta\~{n}o \cite{Aquino1998} and recently used by Jiao et al  \cite{Jiao2017} in the three-dimensional case. We have obtained that the total wavefunction $\Psi_{n,m}^{(r_0)}(\vec r;\alpha)$ of the system can be expressed as 
\begin{equation}\label{confined_wf}
\Psi_{n,m}^{(r_0)}(\vec r;\alpha)=R_{n,m}^{(r_0)}(r;\alpha) \,\phi(\theta),
\end{equation}
where $\alpha$ is a variational parameter and $R_{n,m}^{(r_0)}(r;\alpha)$ denotes the approximate radial wavefunction given by 
\begin{equation}\label{confined_radial_wf}
R_{n,m}^{(r_0)}(r;\alpha)=R_{n,m}(r;\alpha)\,\chi^{(r_0)}(r)
\end{equation}
with the  cut-off function $\chi^{(r_0)}(r)=\left(1-\frac{r}{r_0}\right)$, to take into account the Dirichlet boundary condition at $r= r_0$, and $R_{n,m}(r;\alpha)$ has the form
\begin{equation}
R_{n,m}(r;\alpha)=N'_{nm}(\alpha)e^{-\alpha r} \left( \alpha r\right)^{\lvert m\rvert}L_{n-\lvert m\rvert-1}^{2\lvert m\rvert}\left(\alpha r\right), 
\end{equation}
where the normalization constant $N'_{nm}(\alpha)$ has an expression similar to Eq. \eqref{constant}, and the optimized values of $\alpha$ are variationally derived. For further details see \cite{Aquino1998,Jiao2017,Estanon2019}; the role of the cut-off function has been recently studied in a careful and systematic way for this system \cite{Rojas2019}.

To compute the wavefunctions in momentum space of the two-dimensional confined hydrogen atom we have performed the Fourier integral transform of the position wavefunctions given by the expression \eqref{confined_wf}, obtaining 
\begin{equation}
\begin{split}
\nonumber
\Phi^{(r_{0})}_{n,m}(\vec p)&=\frac{1}{2\pi}\int_{\mathbb R^2}\Psi^{(r_{0})}_{n,m}(r;\alpha)e^{-i\vec{p}\cdot\vec{r}} \text{d} \vec r\\
&=(2\pi)^{-\frac32}\int_{0}^{\infty} R^{(r_{0})}_{n,m}(r;\alpha)\, r\, \left[\int_{0}^{2\pi} e^{im\theta} e^{-i\vec{p}\cdot\vec{r}}\text{d}\theta\right]\,\text{d}r,
\end{split}
\end{equation}
which can be rewritten as
\begin{equation}\label{momentum_wf}
\Phi^{(r_{0})}_{n,m}(\vec p)=\frac{i^{3m}\,e^{im\theta_p}}{(2\pi)^{1/2}}\int_{0}^{r_{0}}R^{(r_{0})}_{n,m}(r;\alpha)\,J_m(r\,p)\, r\text{d}r,
\end{equation}
where we have taken into account the following properties of the Bessel functions: $J_{n}(z)=\frac1{2\pi \,i^n}\int_0^{2\pi}e^{i\,n\tau}e^{i\,z\,cos\tau}\,d\tau$ and $J_n(-z)=(-1)^nJ_n(z)$.\\

\subsection{Probability Densities}

Finally, the corresponding position and momentum probability densities of this two-dimensional system are given by
\begin{equation}\label{posden}
\rho(\vec{r}) \equiv \rho_{n,m}(\vec r;r_0) = \left|\Psi_{n,m}^{(r_0)}(\vec r;\alpha)\right|^2	
\end{equation}
and 
\begin{equation}\label{momden}
 \gamma(\vec{p}) \equiv \gamma_{n,m}(\vec p;r_0) = \left|\Phi_{n,m}^{(r_0)}(\vec p;\alpha)\right|^2	
\end{equation}
which are the basic variables of the information theory of the two-dimensional confined hydrogenic system. They will be used to compute the entropy- and complexity-like quantities of the system in the next sections.\\

\section{Entropic measures}\label{Entropic}
In this section we investigate the confinement dependence of the Shannon entropy and Fisher information for the 1s, 2s, 2p and 3d states of the two-dimensional confined hydrogenic atom in both position and momentum spaces. The corresponding entropic uncertainty relations are also studied. 

Although the entropic measures do not depend on the energies but on the eigenfunctions, let us first comment for completeness that the optimal energies for the aforementioned states of the system have also been obtained in variational form, for which it is necessary to find the minimum value of the functional $E(\alpha)=\langle\Psi|H|\Psi\rangle$ with respect to the variational parameter $\alpha$. The corresponding values $E_{10}, E_{20}, E_{21}$ and $E_{32}$ are given in Table I as a function of the confinement radius $r_0$. We observe that for large values of $r_0$ these energies decrease monotonically towards to their respective values for the free case (free of any confinement). Moreover, the energetic lines $E_{20}$ and $E_{32}$ cross at $r_{0}^{*}\simeq 1$ a.u., giving rise to the so-called "s-d" inversion; this is a common feature in systems confined by cavities with impenetrable walls.
\begin{table}[H]
\caption{Orbital energies $E_{10}, E_{20}, E_{21}$ and $E_{32}$ of the two-dimensional hydrogen atom as a function of the confinement radius $r_0$. Atomic units are used.
}
\centering
 \begin{tabular}{ c c c c c } 
\hline
 $r_0$ & $E_{10}$ & $E_{21}$ & $E_{20}$ & $E_{32}$\\ [0.5ex] 
 \hline\hline
0.50000   & 3.92586   & 25.48515  & 56.01640  & 49.80890   \\   
0.60000   & 1.53084   & 17.10328  & 37.37040  & 34.09951   \\
0.70000   & 0.21363   & 12.12701  & 26.31850  & 24.69224   \\
0.80000   & -0.56133  & 8.94756   & 19.27500  & 18.62865   \\
0.90000   & -1.03983  & 6.80218   & 14.60420  & 14.50030   \\
1.00000   & -1.34601  & 5.29221   & 11.33490  & 11.56791   \\
1.30000   & -1.77456  & 2.74203   & 5.81833   & 6.52929    \\
1.60000   & -1.91438  & 1.54967   & 3.28143   & 4.10112    \\
1.80000   & -1.95294  & 1.08564   & 2.31032   & 3.12980    \\
2.00000   & -1.97315  & 0.76587   & 1.65148   & 2.44530    \\  
3.00000   & -1.99719  & 0.08023   & 0.30176   & 0.88537    \\  
4.00000   & -1.99939  & -0.11005  & -0.03841  & 0.38292    \\  
6.00000   & -1.99991  & -0.20310  & -0.19213  & 0.06512    \\  
8.00000   & -1.99997  & -0.21842  & -0.21631  & -0.02496   \\  
10.00000  & -1.99999  & -0.22129  & -0.22080  & -0.05741   \\  
20.00000  & -1.99999  & -0.22220  & -0.22219  & -0.07961   \\  
\vdots&	\vdots&	\vdots & \vdots & 	\vdots\\
$\infty$  & -2.00000  & -0.22222  & -0.22222  & -0.08000   \\[1ex] 
\hline
\end{tabular}
\label{tablepert}
\end{table}

\subsection{Shannon entropy}
The Shannon entropies of the two-dimensional confined hydrogenic atom with radius $r_{0}$, which is characterized by the position and momentum probability densities $\rho(\vec r)$ and $\gamma(\vec p)$ defined by Eqs. \eqref{posden} and \eqref{momden} respectively, are given \cite{Shannon1948} (see also \cite{Dehesa2011}) by
\begin{equation}\label{posShannon}
S_\rho(r_0)=-\int_{\mathbb R^2} \rho(\vec r;r_0) \log[\rho(\vec r;r_0)]\,d\vec r
\end{equation}
and
\begin{equation}\label{momShannon}
S_\gamma(r_0)=-\int_{\mathbb R^2} \gamma(\vec p;r_0) \log[\gamma(\vec p;r_0)]\,d\vec p,
\end{equation}
respectively. These two position and momentum global quantities are known to satisfy the entropic uncertainty relation \cite{Bialynicki1975,Beckner1975,Rudnicki2012}
\begin{equation}\label{Shannon-ineq}
S_\rho+S_\gamma\ge 2\,\log(e\,\pi) \simeq 4.29 
\end{equation}

\begin{figure}[H]\centering
	\begin{minipage}{0.4\linewidth}
		\includegraphics[width=\linewidth]{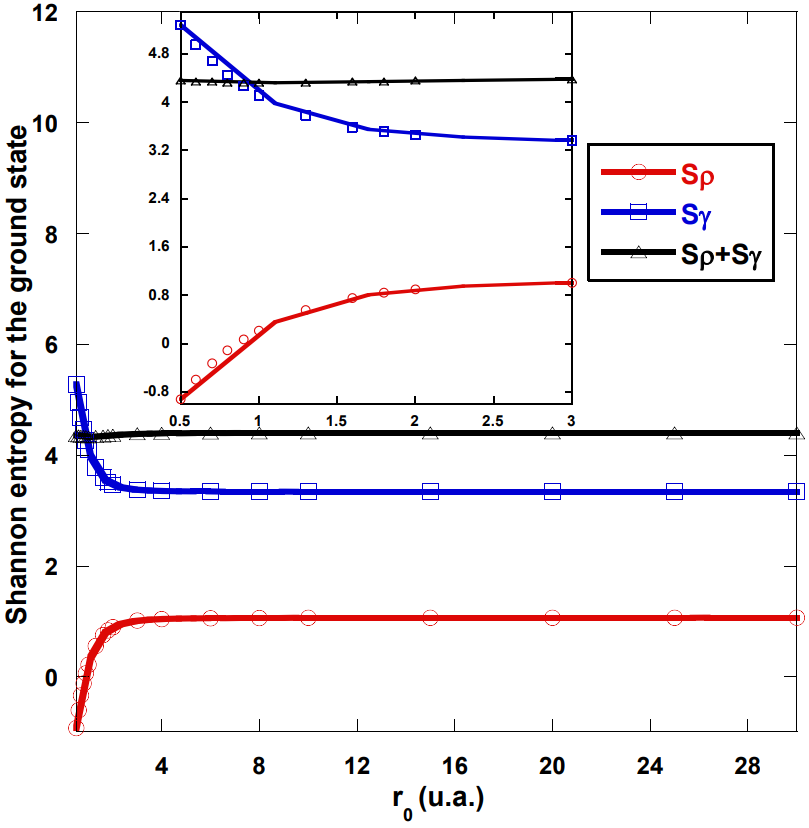}
	\end{minipage}
	\begin{minipage}{0.4\linewidth}
		\includegraphics[width=\linewidth]{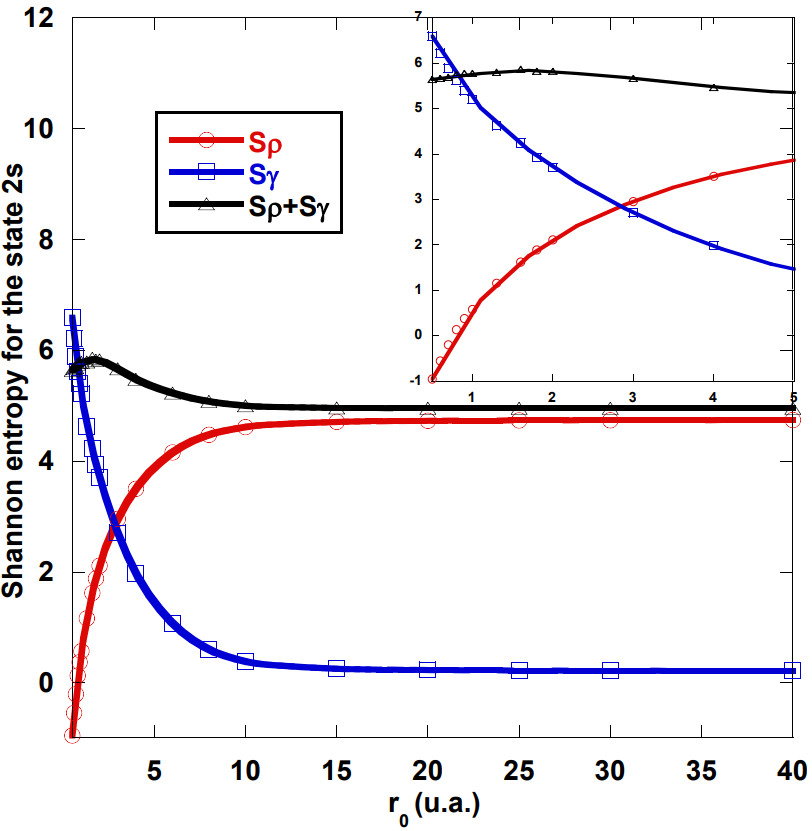}
	\end{minipage}
	\begin{minipage}{0.4\linewidth}
		\includegraphics[width=\linewidth]{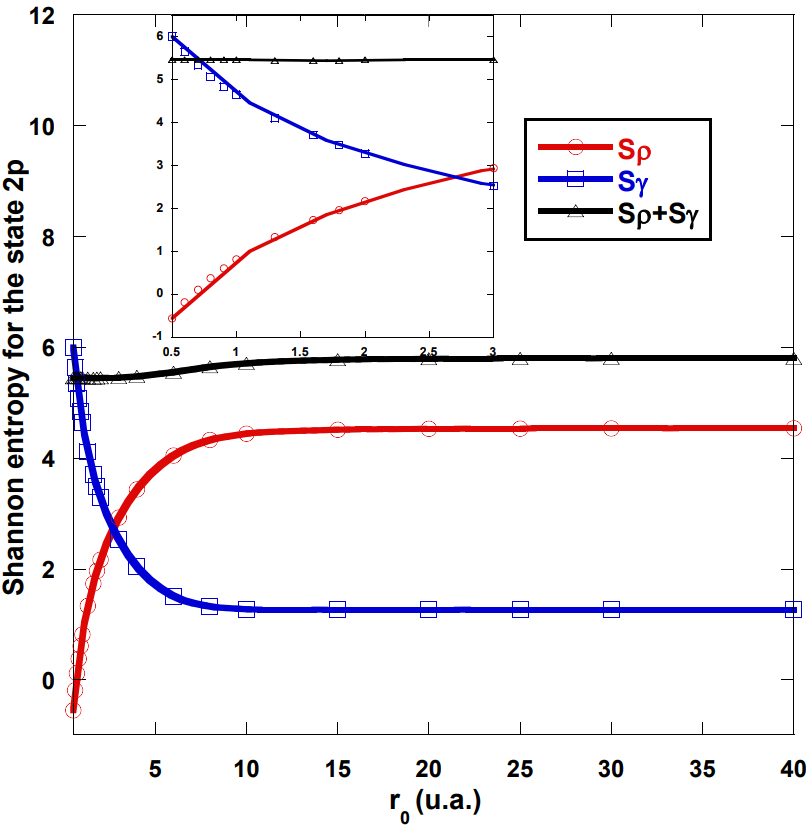}
	\end{minipage}
	\begin{minipage}{0.4\linewidth}
		\includegraphics[width=\linewidth]{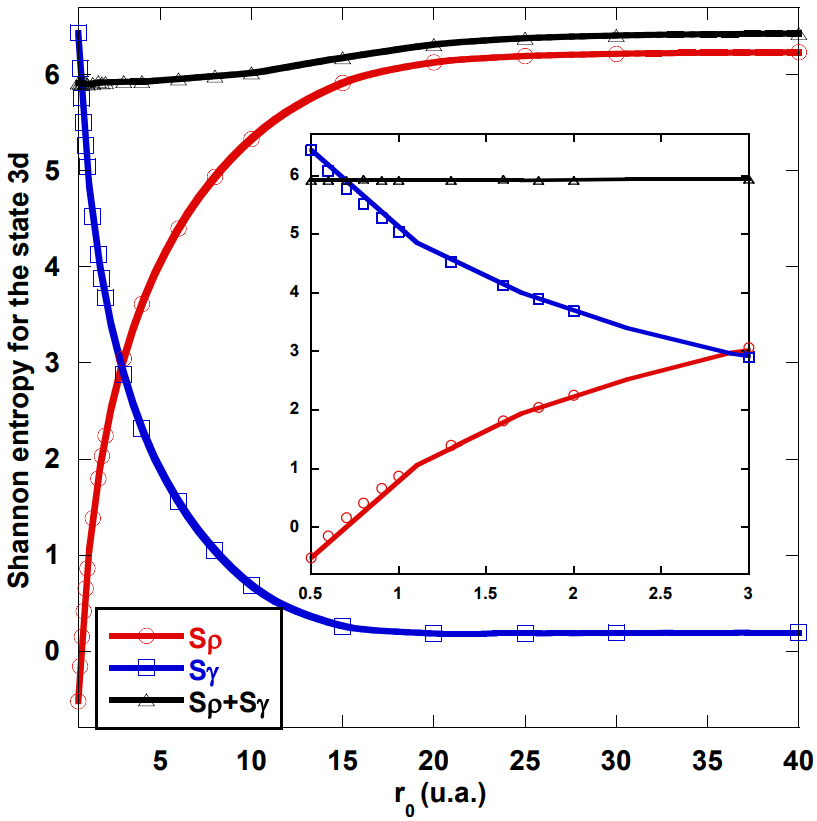}
	\end{minipage}
	\caption{Confinement dependence of the Shannon entropy for the 1s, 2s, 2p and 3d states of the 2D-CHA in both position and momentum spaces. The corresponding uncertainty sum is also given.}
	\label{Fig1a}
\end{figure}

\begin{figure}[H]\centering
	\begin{minipage}{1\linewidth}
		\includegraphics[width=\linewidth]{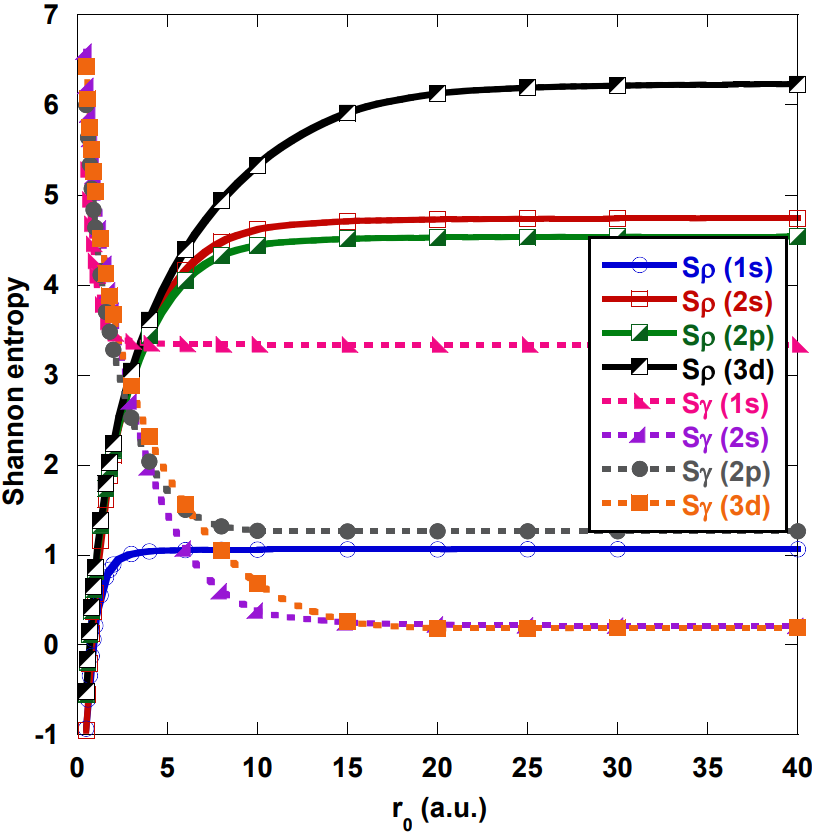}
	\end{minipage}
	\caption{Comparison of the (1s, 2s, 2p, 3d) Shannon entropies of the 2D-CHA in both position and momentum spaces.}
	\label{Fig1b}
\end{figure}

In Figure \ref{Fig1a} we show the Shannon entropy in both position and momentum spaces as a function of the confinement radius $r_0$ for the quantum states 1s, 2s, 2p and 3d.
First we observe that both position ($S_\rho$) and momentum ($S_\gamma$) confined Shannon-entropy values go down to the (exactly known \cite{Dehesa2010}) free values when the confinement radius $r_0$ increases, reaching the latter ones at around $6 \,a.u. (1s), 12\, a.u. (2p), 15\, a.u. (2s)$ and $22\, a.u.$ (3d). This indicates e.g. that for the ground state Shannon entropy gradually decreases as $r_0$ is decreasing from $6 \,a.u.$, as one would expect since the electron density gets more compressed and then the system is more localized; this qualitative behavior was already observed by Sen \cite{Sen2005} in the \textit{three}-dimensional confined hydrogen. The values of the Shannon entropy for the $D$-dimensional free (i.e., unconfined) hydrogen atom have been analytically evaluated by Dehesa et al \cite{Dehesa2010,Toranzo2019}; for example, the ground-state two-dimensional free value is equal to $2+log(\pi/8)=1.06$ (see Eq. (57) in \cite{Dehesa2010}).\\
The behavior of the Shannon entropy in momentum space is reciprocal with respect to the position one, so that the position-momentum sum ($S_\rho+S_\gamma$) gets satisfied according to the entropic uncertainty relation given by Eq. (\ref{Shannon-ineq}).\\ 
Moreover, this global behavior has a number of confinement pecularities for the ground and excited states. They can be also observed in Figure \ref{Fig1b} to better distinguish the ordering of Shannon entropy for the four states at a given radius, and consequently the extent of their electron delocalization.

For the ground state, the Shannon entropy increases (position) and decreases (momentum) monotonically as $r_0$ is increasing; this phenomenon indicates that the electronic cloud is more and more delocalized in configuration space, and more and more concentrated in momentum space, until to reach the free situation. Then, the greater the confinement (i.e.,the smaller $r_0$), the smaller the position Shannon entropy and the greater the momentum Shannon entropy. For the excited states, the Shannon entropy follows a monotonic behavior similar to the ground-state one but only until a critical value $r_c \simeq 2.8\,a.u$; then, the position and momentum entropies intersect, exchanging their mutual behavior. This cross-over phenomenon can be easily seen in the three corresponding graphs, being most apparent in the three small windows covering the  confinement regions $0-3\, a.u.$ and $0-5 \,a.u.$ within these graphs.\\
It is interesting to remark that this cross-over confinement phenomenon, which is not present for the 2D-CHA ground state, does  occur at the ground state in the \textit{three}-dimensional confined hydrogen system, as was pointed out by Sen and others \cite{Sen2005,Aquino2013,Nascimento2018,Mukherjee2018,Mukherjee2018a,Garza2019} although its existence is not explicitly discussed for excited states.\\
Finally, note also that for a given value of the confinement radius the position (momentum) Shannon entropy of the 2D-CHA system grows (decreases) with the energy for all the circular states, but such is not the case for the $(2s)$-state.

\subsection{Fisher information}

The Fisher information for the stationary states of the two-dimensional confined hydrogenic atom, which are characterized by the position probability density $\rho(\vec r;r_0)$ defined by Eq. \eqref{posden}, is given \cite{Fisher1925} (see also \cite{Dehesa2011}) by
\begin{equation}
F_\rho(r_0)=\int_{\mathbb R^2}\frac{\left|\nabla\rho(\vec r;r_0)\right|^2}{\rho(\vec r;r_0)}\,d\vec r
\end{equation}
Similarly, the momentum Fisher information of this system is given by
\begin{equation}
F_\gamma(r_0)=\int_{\mathbb R^2}\frac{\left|\nabla\gamma(\vec p;r_0)\right|^2}{\gamma(\vec p;r_0)}\,d\vec p
\end{equation}
in terms of the momentum density $\gamma(\vec p;r_0)$ defined by Eq. \eqref{momden}.
These two position and momentum local quantities satisfy the uncertainty relation
\begin{equation} \label{Fisher-ineq}
F_\rho\times F_\gamma\ge 4\times 2^2 = 16,
\end{equation} 
for the real wavefunctions of the system \cite{Sanchez-Moreno2011}, what in our case occurs for the $1s$ and $2s$ states only.

\begin{figure}[H]\centering
	\begin{minipage}{0.4\linewidth}
		\includegraphics[width=\linewidth]{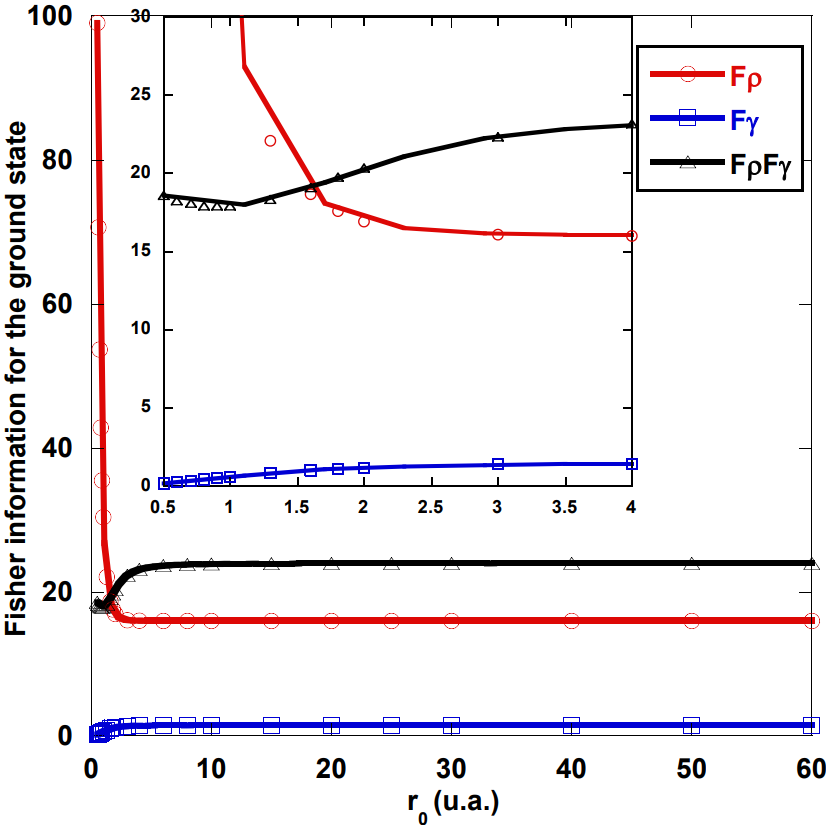}
	\end{minipage}
	\begin{minipage}{0.4\linewidth}
		\includegraphics[width=\linewidth]{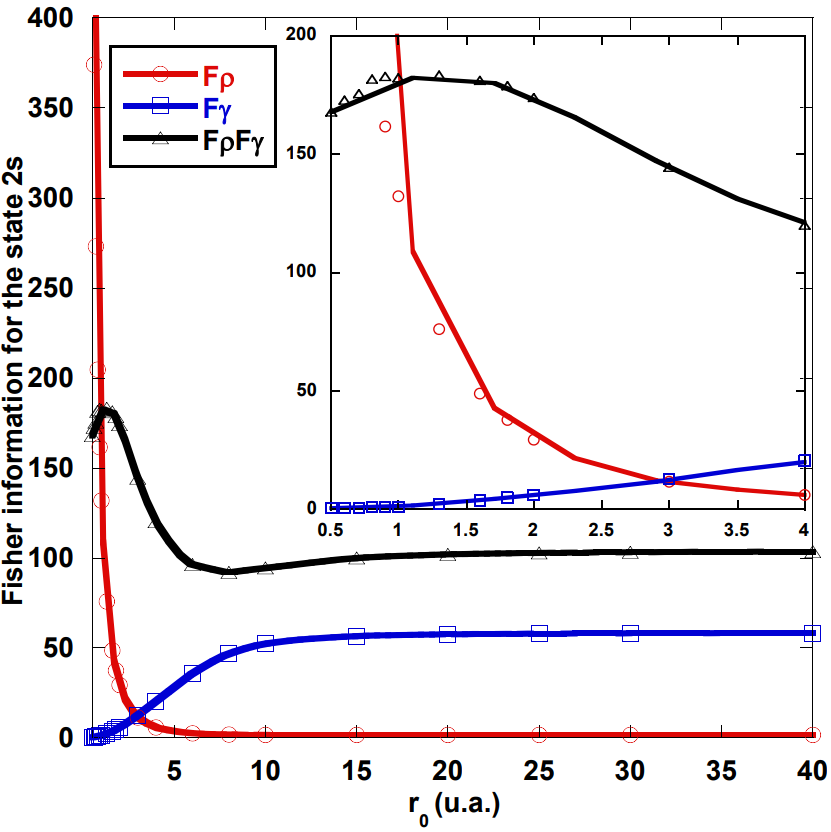}
	\end{minipage}
	\begin{minipage}{0.4\linewidth}
		\includegraphics[width=\linewidth]{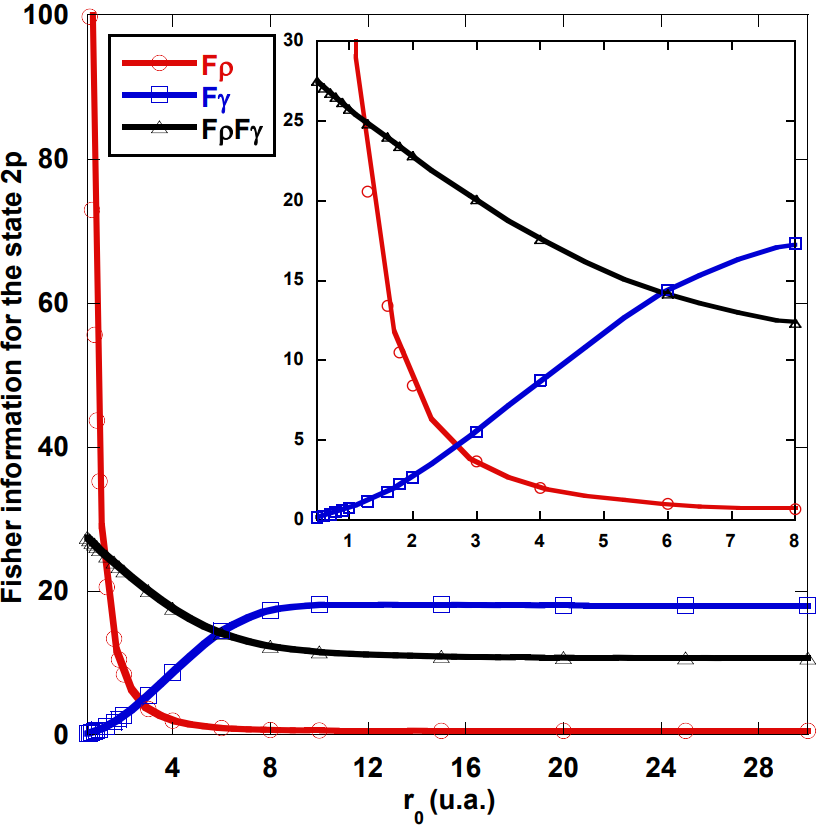}
	\end{minipage}
	\begin{minipage}{0.4\linewidth}
		\includegraphics[width=\linewidth]{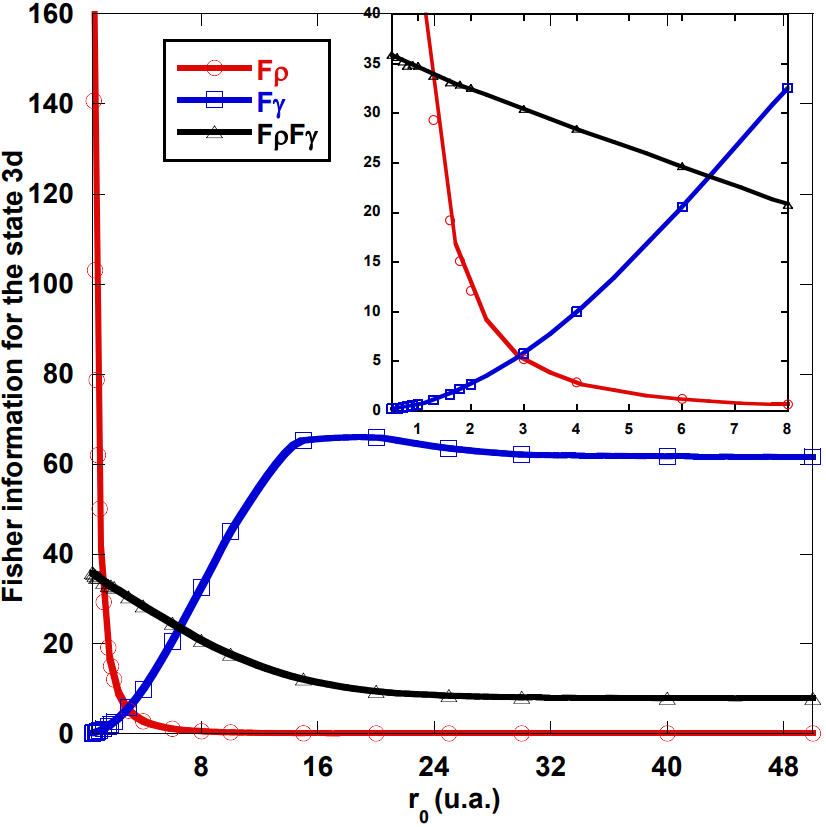}
	\end{minipage}
	\caption{Confinement dependence of Fisher information for the 1s, 2s, 2p and 3d states of the 2D-CHA in both position and momentum spaces. The corresponding uncertainty product is also given.}
	\label{FigFish}
\end{figure}

In Figure \ref{FigFish} we show the Fisher information of our system in both position and momentum spaces as a function of the confinement radius $r_0$ for the quantum states 1s, 2s, 2p and 3d.
Here again we observe the two confinement effects detected by the Shannon entropy, but differently quantified by the Fisher information which is extremely sensitive to the density oscillations of the system. Here however, contrary to the Shannon entropy, the Fisher information decreases (position) and increases (momentum) when $r_0$ is increasing, so that they tend broadly and fastly to the free values in such a way that the Fisher-information-based uncertainty relation (\ref{Fisher-ineq}) for the $1s$ and $2s$ states is always fulfilled because they have real wavefunctions. Keep in mind that the entropic uncertainty relation (\ref{Fisher-ineq}) is only valid for real wavefunctions \cite{Sanchez-Moreno2011}, and the wavefunctions of the $2p$ and $3d$ states have an imaginary character. Note also that the free position and momentum values are reached for values of the confinement radius slightly lower than the ones previously found in the Shannon entropy. Moreover, for completeness, let us remark that the behavior of the position and momentum Fisher information measures of the 2D-CHA is qualitatively similar to the corresponding ones of the 3D-CHA recently investigated for the ground and excited states \cite{Aquino2013,Mukherjee2018b}. For completeness let us mention here that the values of the Fisher information for the $D$-dimensional free (i.e., unconfined) hydrogen atom have been analytically evaluated by Dehesa et al \cite{Dehesa2010}; for example, the two-dimensional free expression for Fisher information of the $(ns)$ hydrogenic states is equal to $\frac{16}{(2n-1)^2}$ (see Eq. (73) in \cite{Dehesa2010}), which gives the values $16$ and $1.78$ for the states $1s$ and $2s$, respectively. \\

Finally, the Fisher-information measures also show up the same cross-over confinement phenomenon for the excited states previously found with the Shannon entropies and at the same critical confinement radius. For the ground state, however, the position and momentum lines do not cross each other, having a monotonic behavior of increasing and decreasing character, respectively. The latter indicates that the electronic charge gets more and more concentrated in the configuration space as the confinement radius is decreasing, basically because the position Fisher information is bigger.\\

\section{Complexity measures}\label{Complexity}

In this section we investigate the confinement dependence of the Fisher-Shannon, LMC and LMC-Rényi complexity measures for the 1s, 2s, 2p and 3d states of the two-dimensional confined hydrogenic atom in both position and momentum spaces. These intrinsic statistical complexities quantify the degree of structure or pattern of the stationary states of the system far beyond the entropic measures.  \\

The Fisher-Shannon complexity measure for a $D$-dimensional probability density $\rho$ is defined as \cite{Romera2004,Angulo2008,Vignat2003}
\begin{equation}
C_{FS}[\rho]=\frac1{2\pi e}F_\rho\times e^{\frac 2D\,S_\rho},
\end{equation}
which quantifies the gradient content (i.e., the oscillatory degree) jointly with the total spreading of $\rho$. So, it is a statistical complexity of local-global character.\\

The LMC-Rényi complexity measure for a $D$-dimensional probability density $\rho$ is defined \cite{Lopez-Ruiz2009,Sanchez-Moreno2014} as 
\begin{equation} \label{LMCRdef}
C_{\lambda,\beta}[\rho]=e^{R_\lambda[\rho]-R_{\beta}[\rho]}, \quad\lambda<\beta, 
\end{equation}
where $R_\lambda[\rho]$ denotes the $\lambda_\textit{th}$-order Rényi entropy of $\rho$ defined as
\begin{equation}
R_\lambda[\rho]=\frac1{1-\lambda} \log\left(\int_{\mathbb R^D}[\rho(\vec r)]^\lambda\,d\vec r\right), \quad \lambda>0
\end{equation}
(which includes the Shannon entropy in the limit $\lambda\to1$). These entropies, which completely characterize the density under certain conditions, quantify various spreading-like facets (governed by the parameter $\lambda$) of the probability density $\rho$. The parameter $\lambda$ has different meanings depending on the context; for instance, it can be interpreted as the inverse of the temperature in thermodynamic systems and it is related to the Reynolds number in turbulence theory. Then, the LMC-Rényi complexity measures $C_{\lambda,\beta}[\rho]$ given by \eqref{LMCRdef} quantify the combined balance of two global spreading aspects of the density $\rho$ with orders $\lambda$ and $\beta$. So, they are statistical complexities of global-global character. \\ 

Note that the case $\lambda=1$ and $\beta=2$ corresponds with the plain LMC (López-Ruiz-Mancini-Calvet) complexity measure given \cite{Catalan2002} by 
\begin{equation}
C_{LMC}[\rho]=e^{S_\rho} \times \mathcal D_\rho \equiv e^{S_\rho}\times\int_{\mathbb R^D}[\rho(\vec r)]^2\,d\vec r,
\end{equation}
where $\mathcal D_\rho = e^{-R_2[\rho]}$ denotes the disequilibrium (also called Onicescu energy) of the $D$-dimensional  system, which quantifies the distance of $\rho$ from equiprobability \cite{Chatzisavvas2005,Alipour2012}. This complexity measure quantifies simultaneously the average height of $\rho(\vec{r})$ (by means of the disequilibrium) and its total extent over the density support (by means of the Shannon quantity).\\

These complexity measures are known to be dimensionless, invariant under translation and scaling transformation \cite{Yamano2004,Yamano2004b}, and universally bounded from below \cite{Guerrero2011,Lopez-Rosa2009} as
\begin{equation}
\label{combounds}
C_{FS}[\rho] \geq D,\,\,C_{LMC}[\rho] \geq 1\,
\text{and}\, C_{\lambda,\beta}[\rho] \geq 1,\, \lambda <\beta 
\end{equation}
for $D$-dimensional probability densities. The corresponding complexity measures for the momentum-space probability density $\gamma(\vec{p})$ will be denoted by $C_{FS} \left[ \gamma \right], C_{LMC} \left[ \gamma \right]$ and $C_{\lambda,\beta} \left[ \gamma \right]$, respectively.\\

In the following we give the main results we have found about the position and momentum complexity measures of the 2D-CHA which are given in Figures \ref{FigComp} and \ref{FigComp2}, respectively. 

\subsubsection*{Position space}

\begin{figure}[H]\centering
	\begin{minipage}{0.3\linewidth}
		\includegraphics[width=\linewidth]{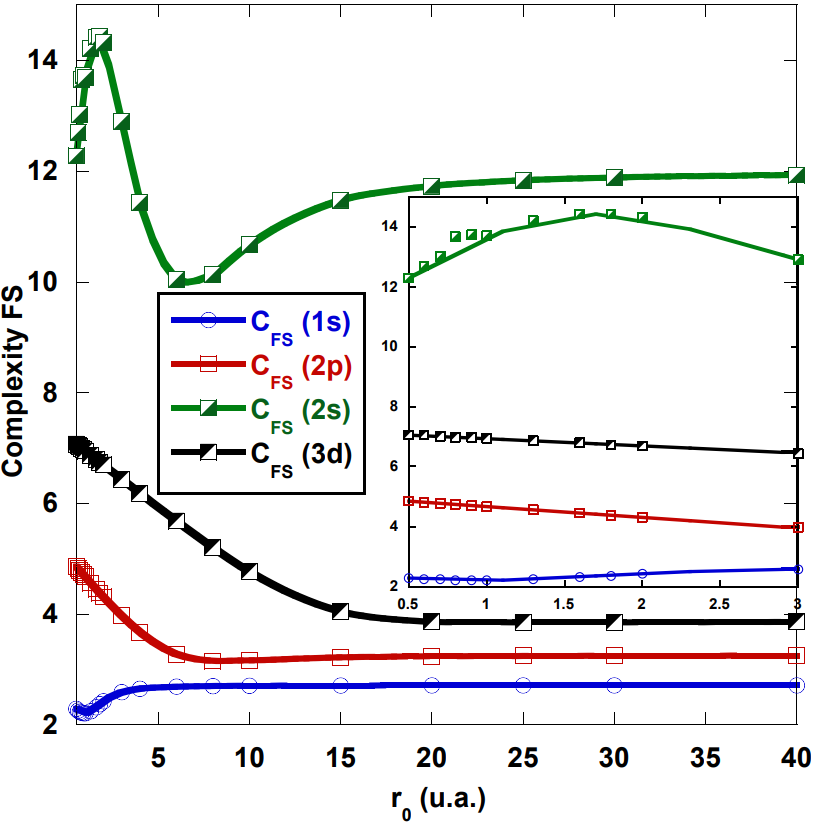}
	\end{minipage}
	\begin{minipage}{0.3\linewidth}
		\includegraphics[width=\linewidth]{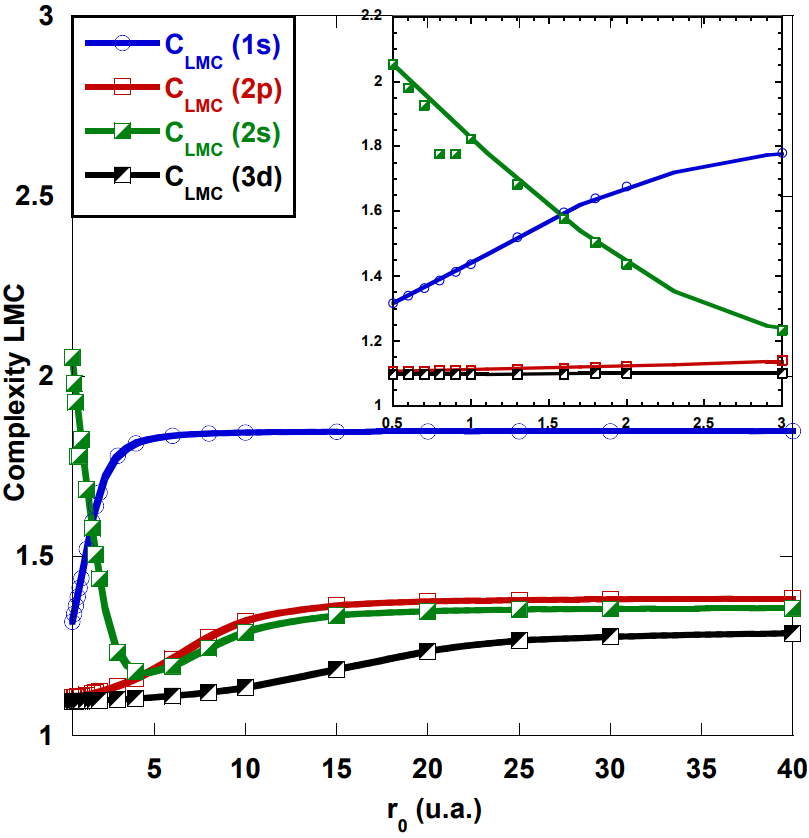}
	\end{minipage}
	\begin{minipage}{0.3\linewidth}
		\includegraphics[width=\linewidth]{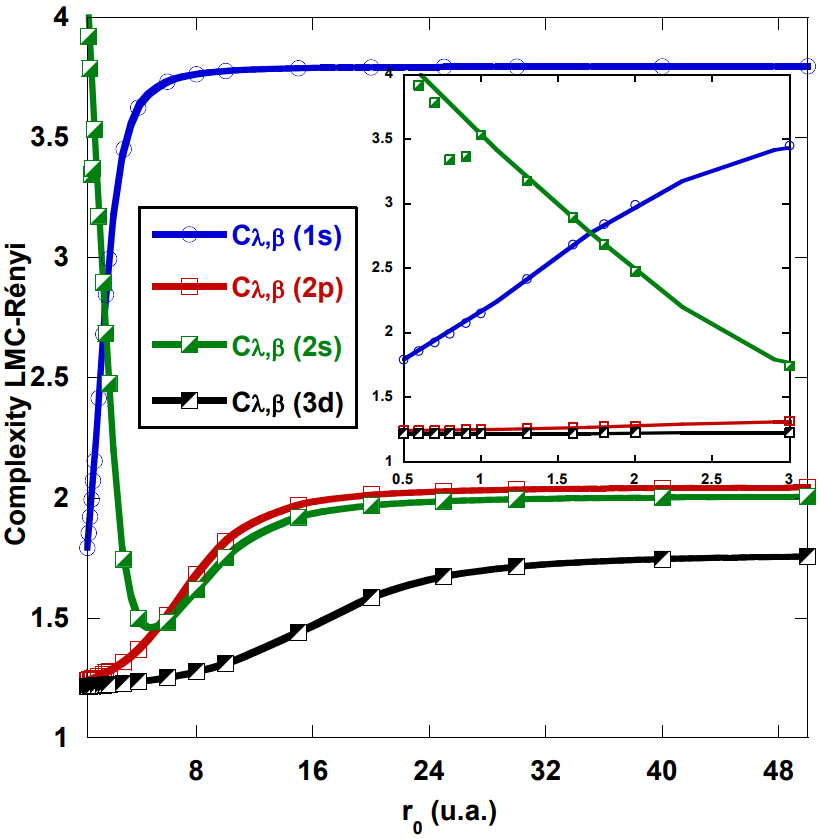}
	\end{minipage}
	\caption{Confinement dependence of the position Fisher-Shannon (\textit{left}), LMC (\textit{center}) and LMC-Rényi (\textit{right}) complexity measures for the 1s, 2s, 2p and 3d states of the 2D-CHA. These quantities are denoted by $C_{FS}(r_0), C_{LMC}(r_0)$ and $C_{\lambda,\beta}(r_0)$ with $\lambda=2/3, \beta=3$, respectively.} 

	\label{FigComp}
\end{figure}

In Figure \ref{FigComp} we show the position Fisher-Shannon $C_{FS}(r_0)$, LMC $C_{LMC}(r_0)$ and LMC-Rényi $C_{\lambda=\frac23,\beta=3}(r_0)$ complexity measures of the two-dimensional confined hydrogen atom as a function of the confinement radius $r_0$ for the quantum states $1s, 2s, 2p$ and $3d$.

We first observe that (a) confinement does distinguish complexity for all stationary states, contrary to what happens in other contexts \cite{Sato2019}; and (b) the three complexity measures fulfil the rigorous lower bounds given by \eqref{combounds}. \\

The three previously defined complexity quantities of the 2D-CHA tend towards to the corresponding constant values of the free (unconfined) hydrogen system for all the ground and excited states when the confinement radius $r_0$ increases. This constancy is reached at $20\, a.u.$ or even earlier. This trend to the free constant values is different for each complexity. In the ground state the three complexities increase monotonically when $r_0$ increases up until $8\,a.u.$, where the constancy is reached. In the first excited state ($2s$) the Fisher-Shannon measure behaves differently than the LMC and LMC-Rényi measures when $r_0$ increases, basically because the former one has a local-global character and the two latter ones have global-global characters: while the Fisher-Shannon measure oscillates for strong confinement (i.e., small $r_0$), both LMC and LMC-Rényi measures have a monotone decrement up to a minimum and then it monotonically increases up until the free constant value when $r_0$ is increasing. In the circular states $2p$ and $3d$ the Fisher-Shannon measures decreases smoothly down to the corresponding free constant values, and both LMC and LMC-Rényi measures increases monotonically up until the corresponding free constant values when $r_0$ increases. \\ 

Moreover, Figure \ref{FigComp} also shows the ordering of the three complexity measures to be  $$C_{FS}[1s] < C_{FS}[2p] < C_{FS}[3d] < C_{FS}[2s]$$ for the Fisher-Shannon measure at all confinements and $$C_{LMC}[1s] > C_{LMC}[2p] > C_{LMC}[2s] > C_{LMC}[3d]$$ for the LMC measure (as well as for the LMC-Rényi measure) at $r_0 \gtrapprox 5\,a.u.$. Note also that $$C_{LMC}[1s] > C_{FS}[1s] > C_{\frac23,3}[1s]$$ for the relative values of the considered complexities of the ground state and $$C_{LMC}[x] > C_{\frac23,3}[x] > C_{FS}[x]$$ for the relative values of the complexities of the excited states $[x] = [2p], [2s], [3d]$ at $r_0 \gtrapprox 5\,a.u.$.. The situation gets more involved at stronger confinements. The physical interpretation of these inequalities follows from the nodal structure of the corresponding wavefunctions and the entropic components of the complexity measures. Finally, there are scanty works to compare with, save the recent efforts of Aquino et al \cite{Aquino2013} for the ground state and Majumdar et al \cite{Majumdar2017} for various excited states of the \textit{three}-dimensional hydrogen atom with some complexity measures with different normalizations. Let us just point out that the very different behavior of the complexity measures for circular  and non-circular states of the 2D-CHA is in accordance with the result for the 3D-CHA.


\subsubsection*{Momentum Space}
\begin{figure}[H]\centering
	\begin{minipage}{0.3\linewidth}
		\includegraphics[width=\linewidth]{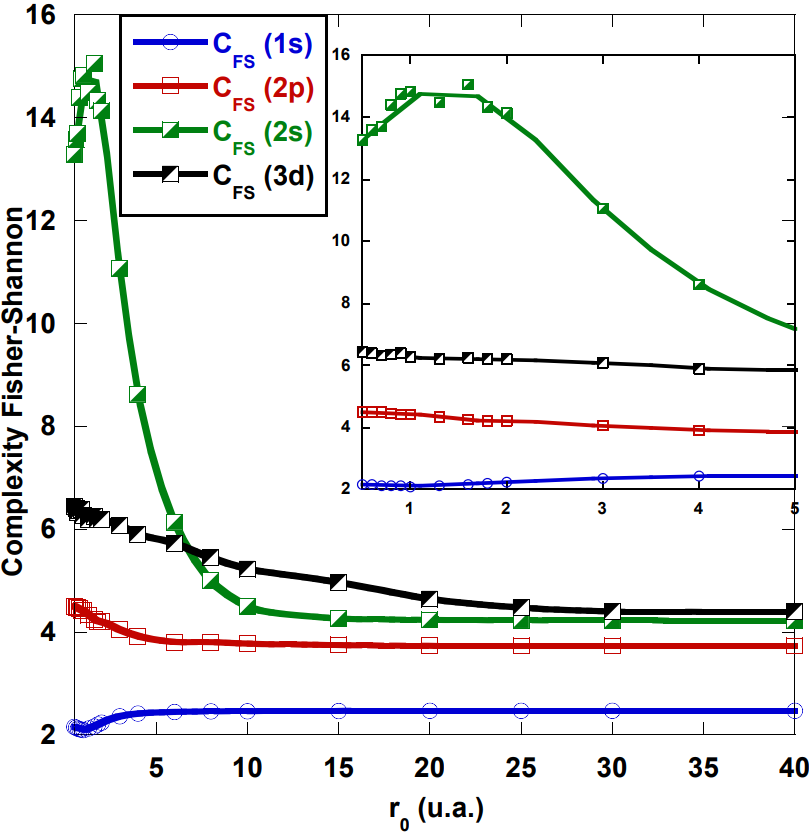}
	\end{minipage}
	\begin{minipage}{0.3\linewidth}
		\includegraphics[width=\linewidth]{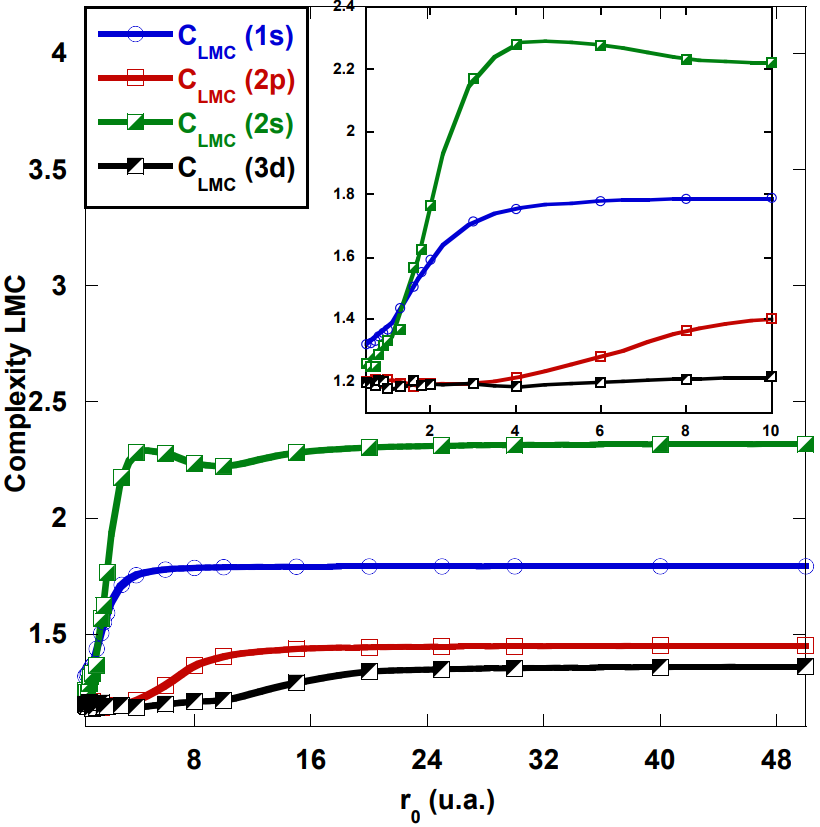}
	\end{minipage}
	\begin{minipage}{0.3\linewidth}
		\includegraphics[width=\linewidth]{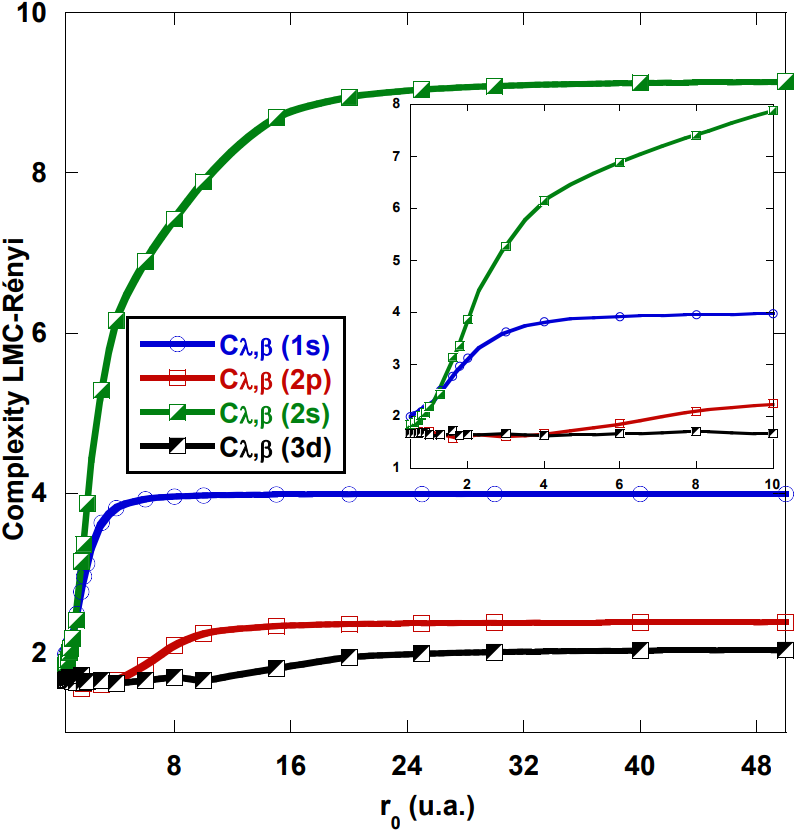}
	\end{minipage}
	\caption{Confinement dependence of the momentum Fisher-Shannon (\textit{left}), LMC (\textit{center}) and LMC-Rényi (\textit{right}) complexity measures for the 1s, 2s, 2p and 3d states of the 2D-CHA. These quantities are denoted by $\mathcal{C}_{FS}(r_0), \mathcal{C}_{LMC}(r_0)$ and $\mathcal{C}_{\lambda,\beta}(r_0)$ with $\lambda=2/3, \beta=3$, respectively.} 	

	\label{FigComp2}
\end{figure}

In Figure \ref{FigComp2} we show the momentum Fisher-Shannon $C_{FS} \left[ \gamma \right](r_0 )\equiv \mathcal{C}_{FS}(r_0)$, LMC $C_{LMC} \left[ \gamma \right](r_0) \equiv \mathcal{C}_{LMC}(r_0)$ and LMC-Rényi $C_{\lambda=\frac23,\beta=3} \left[ \gamma \right](r_0) \equiv \mathcal{C}_{\lambda=\frac23,\beta=3}(r_0)$ complexity measures of the two-dimensional confined hydrogen atom as a function of the confinement radius $r_0$ for the quantum states $1s, 2s, 2p$ and $3d$.

Here again we observe that for all ground and excited states the three momentum complexity measures of the 2D-CHA (a) fulfil the universal bounds \eqref{combounds} and (b) tend towards to the corresponding constant free values when the confinement radius $r_0$ increases. This constancy is reached at $20\, a.u.$ or even earlier, as in position space. This behavior towards free constancy is different for each complexity. In the ground state the three momentum complexities smoothly increase when $r_0$ increases up until the free constant value which is reached at $8\,a.u.$. In the first excited state ($2s$) the confinement dependence of the Fisher-Shannon measure is very different with respect to the one of the LMC and LMC-Rényi measures: the former measure has a pronounced maximum at small $r_0$, between $1$ and $2\, a.u.$, from which it rapidly goes down to the free constancy and the two other complexity measures have an increasing behavior up to the corresponding free constant values. \\

The relative values for the complexity measures of the 2D-CHA can also be extracted from the figure. They are given by the inequalities $$\mathcal{C}_{FS}[1s] < \mathcal{C}_{FS}[2p] < \mathcal{C}_{FS}[2s] < \mathcal{C}_{FS}[3d]$$ for the Fisher-Shannon measure and $$\mathcal{C}_{LMC}[2s] > \mathcal{C}_{LMC}[1s] > \mathcal{C}_{LMC}[2p] > \mathcal{C}_{LMC}[3d]$$ for the LMC measure (as well as for the LMC-Rényi measure) at $r_0 \gtrapprox 5\,a.u.$ Note also that $$\mathcal{C}_{LMC}[x] > \mathcal{C}_{\frac23,3}[x] > \mathcal{C}_{FS}[x]$$ for the states $[x] = [1s], [2s]$ at $r_0 \gtrapprox 1$ and $5\,a.u.$, respectively, and $$\mathcal{C}_{FS}[x] > \mathcal{C}_{\frac23,3}[x] > \mathcal{C}_{LMC}[x]$$ for the states $[x] = [2p], [3d]$ at all confinements. The comparison of these results with other momentum efforts highlights the ability of the statistical complexity measures to capture the nodal structure of the CHA, as they are able to clearly distinguish  the circular and non-circular states in both two-dimensional and three-dimensional systems.  This is because they satisfy a number of mathematical requirements as already pointed out. However, other proposals of complexity-like measures \cite{Majumdar2017}  that do not satisfy  generally the full set of criteria of an intrinsic statistical complexity \cite{Rudnicki2016} show a similar behavior for all states, not being able to fully capture the internal structure of the system.

\section{Conclusions}\label{Conclusions}

In this work we have studied the confinement dependence of the main entropic measures (Shannon, Fisher) and the complexity-like measures (Fisher-Shannon, LMC, LMC-Rényi) for a few stationary states of the two-dimensional hydrogen atom in position and momentum spaces. These one- and two-component information-theoretic measures do not depend on the state's energy, but they are functional integrals of the quantum-mechanical eigenfunction of the system's state. They quantify in a quite complete way various local and global aspects of the internal disorder of the system which are closely connected to the electronic and pointwise concentration of the electronic charge all over the confinement region. \\

We have found that the position and momentum ground-state Shannon entropies increase and decrease, respectively, when the confinement radius $r_0$ is increasing so that (a) they tend without crossing to the corresponding free values in a fast, monotone way, reaching them at around $6\,a.u.$ and (b) the entropic uncertainty relation is always preserved. The ground-state Fisher information has a somewhat reciprocal confinement-dependent behavior, so that again its values tend without crossing to the free values when $r_0$ increases and the associated Fisher-information uncertainty relation is satisfied. Then, the greater the confinement (i.e.,the smaller $r_0$), the smaller the position Shannon entropy and the greater the position Fisher information. Note that ground-state confinement effects are appreciable, quantifiable when $r_0< 6\,a.u.$

For the excited states $2s, 2p$ and $3d$ we have found a cross-over phenomenon at the critical confinement  $r_c \simeq 2.8\,a.u$ in both entropic cases of Shannon and Fisher types. Indeed when $r_0$ increases the position and momentum  entropic measures intersect at $r_c$, exchanging its mutual relationship but always tending smoothly to the corresponding free values.

 We have also shown that confinement does distinguish complexity of the 2D-CHA for all stationary states. This is done by means of the Fisher-Shannon, LMC and LMC-R\'enyi measures. All these quantities tend in slightly different ways towards to the corresponding constant free values for both ground and excited states when the confinement radius $r_0$ increases. This constancy is reached at around $20\, a.u.$. Then, these complexity measures can detect the confinement effects at $r_0 >> 6\,a.u.$, basically due to the nodal structure of the state (2s) and their two entropic components.
 
Finally, for completeness, let us point out that various authors have recently considered various information entropies of global (Shannon, R\'enyi, Tsallis) \cite{Mukherjee2018,Mukherjee2018a,Aquino2013} and local (Fisher, relative Fisher) \cite{Mukherjee2018b,Mukherjee2018c,Aquino2013,Wu2020,Yu2017} character for numerous excited states of \textit{three}-dimensional free and confined hydrogen-like systems and for the helium atom \cite{Ou2019}, discovering a number of interesting delocalization features. The Shannon entropies of various atomic and molecular potentials have also been recently calculated \cite{Najafizade2016a,Najafizade2016b,Najafizade2017}.

\section*{Funding information}
{The work of J.S. Dehesa has been partially supported by the grant FIS2017-89349P of the Agencia Estatal de Investigaci\'on (Spain) and the European Regional Development Fund (FEDER).}

\section*{Author contributions}

N.A., C.R.E. and J.S.D. contributed in conceptualization, data curation, investigation, and writing of the original draft of the manuscript. C.R.E. and D.P.C. performed formal analysis, writing of the review, and edited the manuscript.

\end{document}